\documentclass[10pt,a4paper,twocolumn]{article}

\usepackage[margin=0.75in]{geometry}
\usepackage{graphicx}
\usepackage{amsmath}
\usepackage{cite}
\usepackage{float}

\title{Sweep-Direction-Dependent Level-Attraction- and Level-Repulsion-Like Signatures in a Twisted, Helically Corrugated Microwave Resonator}
\author{E. C. I. Paterson and J. Bourhill\\[0.5em]
  \small Quantum Technologies and Dark Matter Labs, Department of Physics,\\
  \small The University of Western Australia, Crawley, WA 6009, Australia\\[0.3em]
  \small \texttt{emma.paterson@research.uwa.edu.au}}
\date{}

\begin{document}
\maketitle

\begin{abstract}
We investigate photon--photon interactions in a mechanically tunable, helically corrugated rectangular microwave resonator. Mechanical twisting modifies the electromagnetic helicity and resonant frequencies of the supported modes, allowing pairs of electromagnetic modes to be tuned into near-degeneracy. Transmission magnitude and phase measurements reveal two distinct responses: a level-attraction-like response characterised by converging antiresonances and a sharp phase variation appearing as a $2\pi$ discontinuity in the wrapped transmission phase, and a level-repulsion-like response characterised by an anticrossing-like transmission feature. For the mode-interaction regions investigated, reversing the twist sweep direction interchanges these responses. We discuss mechanical hysteresis and changes in electromagnetic backscattering and modal interference as possible origins of this behaviour. These observations suggest a potential route towards microwave devices with mechanically switchable level-attraction-like and level-repulsion-like transmission responses.
\end{abstract}

\section{Introduction}

Coupled resonant systems can exhibit different responses depending on the nature of the interaction. Coherent coupling can produce level repulsion, characterised by an anticrossing of the coupled modes, whereas dissipative coupling can produce level attraction, characterised by the convergence of the coupled modes~\cite{Harder2018,Rao2019}. Controlling attractive and repulsive coupling behaviour offers potential applications in quantum sensing and nonreciprocal microwave devices~\cite{Harder2018}, as well as metrology and RF technologies~\cite{Bourcin2024}.

Our previous work demonstrated that mechanical twist can generate electromagnetic helicity in microwave cavities and enable continuous tuning of their resonant frequencies~\cite{Paterson2025Helicity,Paterson2026}. Here, we exploit this tunability in a resonator with fixed right-handed corrugation to bring pairs of photon modes into near-degeneracy and investigate their resulting interactions through microwave transmission measurements.

We observe distinct level-attraction-like and level-repulsion-like signatures in the transmission spectra, which interchange when the direction of the twist sweep is reversed for the mode-interaction regions investigated. We consider mechanical hysteresis and changes in electromagnetic backscattering and modal interference as possible physical interpretations of the observed switching.

\section{Resonator and Measurements}

\begin{figure}[!b]
\centering
\includegraphics[height=6.5cm]{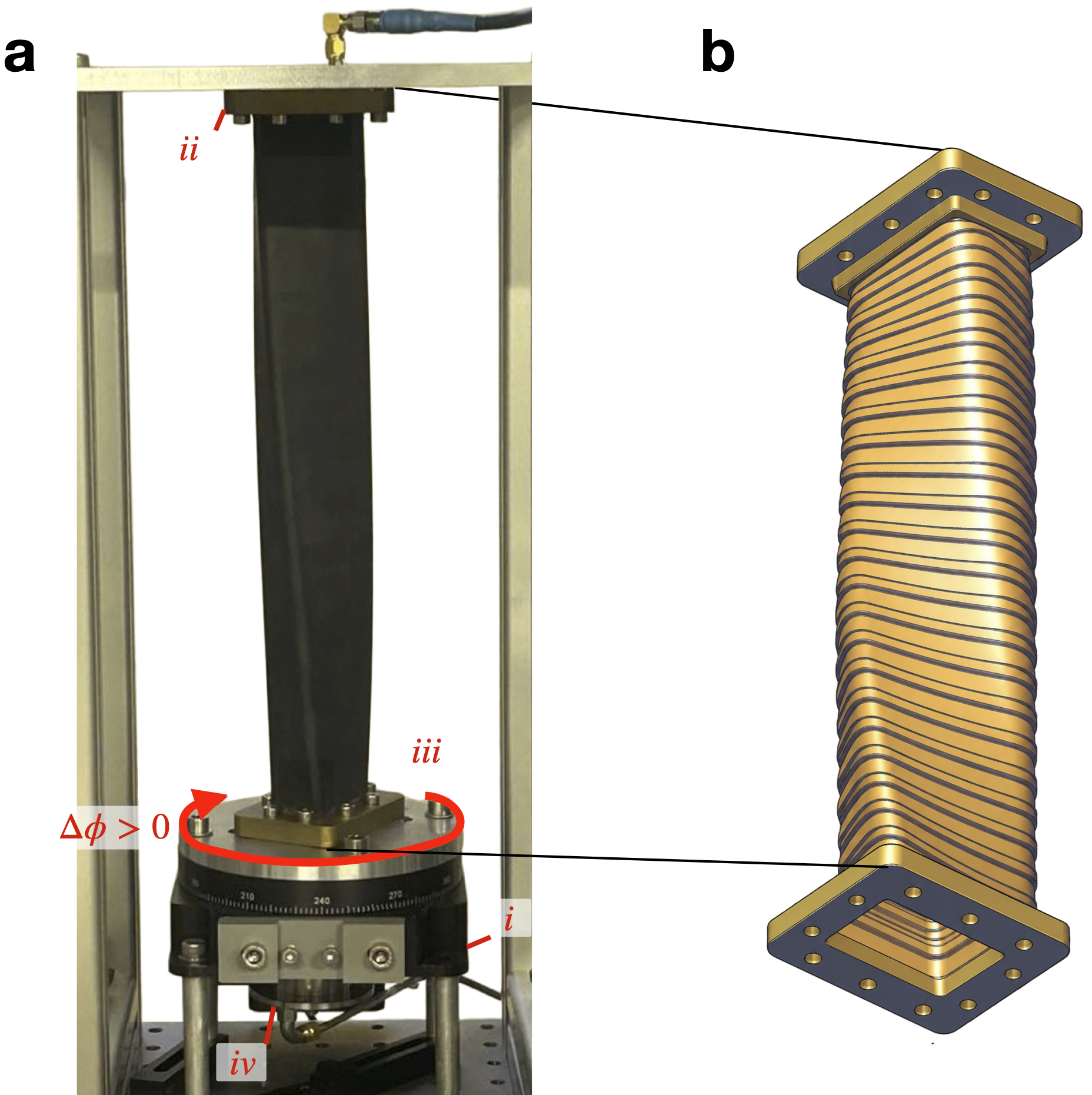}
\caption{Experimental arrangement (a) used to impose a controlled twist on a helically corrugated WR-137 rectangular resonator. Mechanical rotation was applied to one end of the resonator using a rotary stage (i), while the opposite end was held fixed. Excitation and detection were achieved through coaxial probes mounted in the end caps (ii, iii). To eliminate cable torsion during rotation, the lower probe was coupled through a rotating connector (iv). The accompanying conceptual illustration (b) demonstrates the right-handed helical corrugation and does not represent the exact geometry or dimensions of the experimental resonator.}
\label{fig:setup}
\end{figure}

A flexible WR-137 waveguide is terminated by metallic end caps. The internal surface of the waveguide is corrugated, with the corrugations following a right-handed helical path along its length. Figure~\ref{fig:setup} shows the experimental arrangement alongside a conceptual illustration of the right-handed helical corrugation. Its internal cross section is $35.48\times16.1$~mm and its length is $312.3$~mm. One end is fixed and a motorised rotation stage rotates the other about the resonator's central axis, providing quasistatic control of the twist angle, $\phi$. Negative and positive values of $\phi$ denote left- and right-handed twist configurations, respectively, while the sign of $\Delta\phi$ specifies the direction of the twist sweep. A vector network analyser measures the complex forward transmission coefficient $S_{21}$ through coaxial probes. A rotating connector prevents cable torsion and isolators suppress line resonances. We analyse both the transmission magnitude, expressed in decibels as $S_{21,\mathrm{dB}}=20\log_{10}|S_{21}|$, and the transmission phase, $\Phi=\angle S_{21}$.

Mechanical twisting modifies the resonator's electromagnetic boundary conditions and can induce hybridisation between transverse electric (TE) and transverse magnetic (TM) modes, producing hybrid modes with nonzero electromagnetic helicity. This helicity can be quantified through a normalised volume integral of $\operatorname{Im}[\mathbf{E}\cdot\mathbf{H}^{*}]$~\cite{Paterson2025Helicity,Paterson2026}. The accompanying changes in the modal field distributions and resonant frequencies allow distinct pairs of hybridised photon modes to be tuned into near-degeneracy~\cite{Paterson2026}.

\section{Transmission Signatures}

\begin{figure}[H]
\centering
\includegraphics[width=0.9\columnwidth]{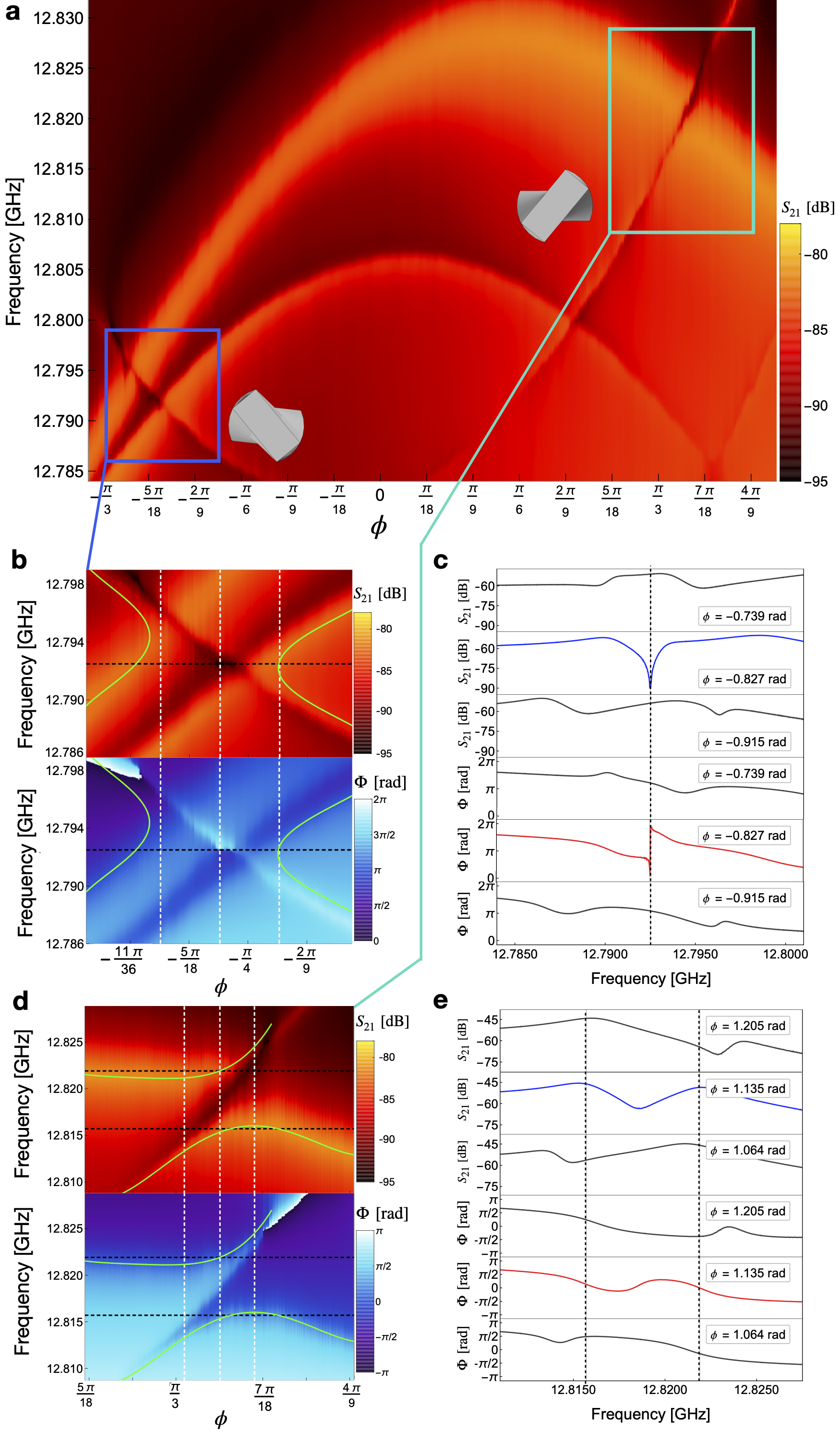}
\caption{Transmission and phase spectra of the rectangular resonator as a function of the applied twist angle, $\phi$, for $\Delta\phi>0$. Panel (a) shows the full transmission contour plot, with the highlighted regions indicating the two mode interaction regions considered in this work. Panels (b) and (d) show zoomed-in transmission (upper) and phase (lower) contour plots for the negative- and positive-twist configurations, respectively. Panels (c) and (e) show representative transmission magnitude, $S_{21}$ (dB), and phase, $\Phi$, spectra at three different twist angles for the corresponding configurations. A level-attraction-like response is observed at $\phi=-0.827$~rad, characterised by a narrow transmission minimum accompanied by a sharp $2\pi$ discontinuity in the wrapped transmission phase, while a level-repulsion-like response is observed at $\phi=1.135$~rad, characterised by separated resonant features and two distinct phase shifts. The vertical white gridlines in panels (b) and (d) indicate the twist angles of the representative spectra in panels (c) and (e), respectively, while the horizontal black gridlines mark the frequencies indicated by the vertical gridlines in the corresponding spectra. The solid green curves on the contour plots trace the resonant transmission peaks associated with the modes being tuned into near-degeneracy.}
\label{fig:spectra}
\end{figure}

Figure~\ref{fig:spectra} compares the transmission magnitude and phase responses near two photon-mode interaction regions as the twist angle is increased ($\Delta\phi>0$). The full transmission contour plot [Fig.~\ref{fig:spectra}(a)] provides an overview of the resonator's spectral response, with the two highlighted regions corresponding to the mode interactions examined in detail in the subsequent panels. For visualisation, the transmission phase is wrapped into $[0,2\pi)$ for the negative-twist configuration and $[-\pi,\pi)$ for the positive-twist configuration, with the same convention applied to the corresponding contour plots and individual spectra. In the negative-twist configuration [Figs.~\ref{fig:spectra}(b),(c)], the transmission spectrum exhibits the convergence of two antiresonances towards a common frequency near $12.7925$~GHz [Fig.~\ref{fig:spectra}(b)]. At $\phi=-0.827$~rad, a narrow transmission minimum is resolved within a broader spectral feature [Fig.~\ref{fig:spectra}(c)], accompanied by a sharp $2\pi$ discontinuity in the wrapped transmission phase, $\Phi$, at approximately the same frequency. Together, the converging antiresonances, narrow transmission minimum, and single phase jump suggest a level-attraction-like transmission response.

Conversely, level-repulsion-like behaviour is observed in the positive-twist configuration, as shown in Figs.~\ref{fig:spectra}(d),(e). The transmission spectrum exhibits two distinct resonant features that approach and then separate in frequency as the twist angle is varied, forming an anticrossing-like feature [Fig.~\ref{fig:spectra}(d)]. At $\phi=1.135$~rad, the two resonant features are resolved in the transmission spectrum [Fig.~\ref{fig:spectra}(e)], accompanied by two distinct phase shifts at the corresponding frequencies. Together, the separated resonant features and corresponding phase shifts suggest a level-repulsion-like transmission response.

\subsection{Sweep-direction-dependent \mbox{switching}}

\begin{figure}[!t]
\centering
\includegraphics[width=1\columnwidth]{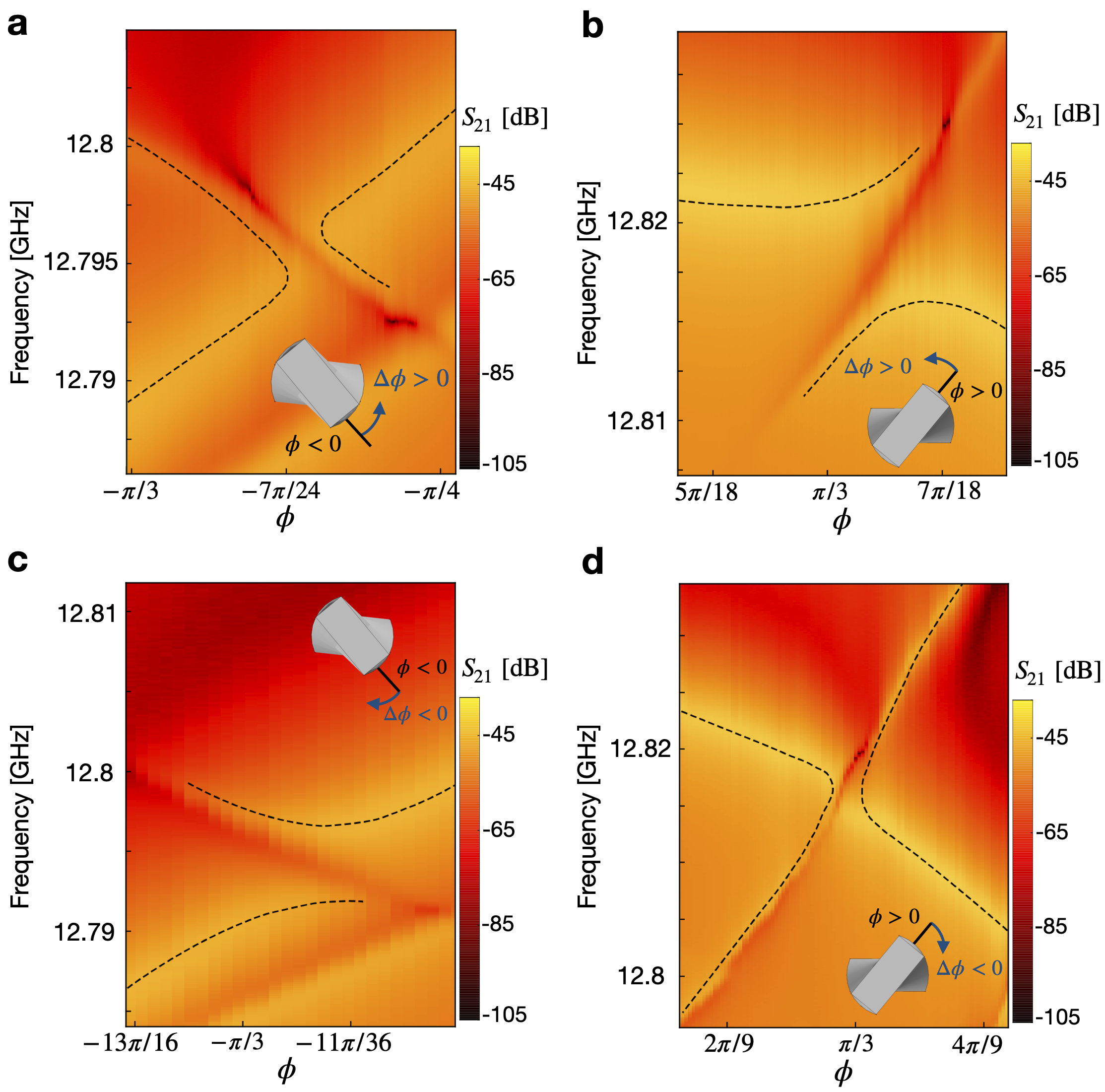}
\caption{Transmission spectra of the rectangular resonator as a function of the applied twist angle, $\phi$, near the mode interaction regions for both sweep directions. Panels (a) and (b) show the spectra for $\Delta\phi>0$, exhibiting level-attraction-like and level-repulsion-like responses, respectively. Panels (c) and (d) show the corresponding spectra for $\Delta\phi<0$, exhibiting level-repulsion-like and level-attraction-like responses, respectively. The dashed black curves on the contour plots trace the resonant transmission peaks associated with the modes being tuned into near-degeneracy.}
\label{fig:sweeps}
\end{figure}

Figure~\ref{fig:sweeps} compares the transmission spectra measured for opposite twist sweep directions, $\Delta\phi>0$ and $\Delta\phi<0$, in two distinct mode-interaction regions occurring at negative and positive twist angles, respectively. For the negative-twist interaction region, a level-attraction-like response is observed when $\Delta\phi>0$ [Fig.~\ref{fig:sweeps}(a)], whereas reversing the sweep direction to $\Delta\phi<0$ produces a level-repulsion-like response [Fig.~\ref{fig:sweeps}(c)]. Conversely, the positive-twist interaction region exhibits a level-repulsion-like response for $\Delta\phi>0$ [Fig.~\ref{fig:sweeps}(b)] and a level-attraction-like response for $\Delta\phi<0$ [Fig.~\ref{fig:sweeps}(d)]. Thus, reversing the direction of the twist sweep interchanges the observed transmission signatures for the two mode interactions investigated. This behaviour is not necessarily universal across all modes supported by the resonator.

\section{Interpretation and Discussion}

The level-attraction-like spectra are consistent with a dissipative component of the effective photon--photon coupling, although the present transmission measurements do not uniquely determine the underlying coupling mechanism. In cavity--magnon systems, dissipative coupling can arise through the cavity Lenz effect, in which magnetisation dynamics induce an RF cavity current whose backaction opposes the magnetisation dynamics~\cite{Harder2018}. Although this specific mechanism does not transfer directly to the present all-photonic system, it illustrates how dissipative interactions can give rise to level attraction. For the twisted resonator investigated here, shared conducting-wall losses, port-mediated coupling, or other common dissipative channels may contribute to an effective dissipative interaction between the photon modes.

The convergence of the antiresonances and the narrow transmission minimum observed in the level-attraction-like response may arise from destructive interference between modal contributions and should therefore not be interpreted, by themselves, as evidence of mode coalescence~\cite{Rao2019,Bourcin2024}. Quantitative modelling of the complex eigenfrequencies and microwave transmission response would be needed to determine the relative coherent and dissipative coupling contributions.

The observed sweep-direction dependence may arise from mechanical hysteresis in the flexible resonator. Increasing and decreasing the applied twist may produce different distributions of mechanical tension and deformation within the corrugated waveguide, even at the same nominal twist angle. These variations in geometry could modify the electromagnetic boundary conditions, modal field distributions, and effective coupling between photon modes. Changes in the geometry of the corrugations may also influence the interference between modal contributions to the transmission response, with variations in reactive loading and backscattering providing possible additional contributions. These effects may contribute to the observed switching, alongside possible changes in conducting-wall losses and probe coupling. Although mechanical twisting modifies electromagnetic helicity and resonant frequencies~\cite{Paterson2025Helicity,Paterson2026}, the role of helicity in the observed photon-mode interaction signatures remains to be established.

The relative contributions of coherent and dissipative coupling and modal interference to the observed transmission signatures also require further investigation. Measurements using resonators with higher quality factors would allow the interacting spectral features and linewidths to be resolved more precisely. Combined with repeated forward and reverse twist sweeps and quantitative modelling of the complex eigenfrequencies and transmission response, these measurements could help establish the role of mechanical hysteresis, modal coupling, and interference in the observed switching. Irrespective of the underlying mechanism, the observed switching between level-attraction-like and level-repulsion-like transmission responses suggests potential applications in mechanically reconfigurable microwave devices, where control over resonant frequencies and associated phase responses could be exploited for sensing, metrology, and frequency-selective RF technologies.

\section{Conclusion}

We observe distinct level-attraction-like and level-repulsion-like transmission signatures in a mechanically tunable corrugated rectangular microwave resonator. For the mode-interaction regions investigated, reversing the direction of the twist sweep interchanges the observed responses. Mechanical hysteresis may contribute to this behaviour through changes in the resonator geometry, modal coupling, and interference between electromagnetic modes, although the underlying mechanism remains to be established. These observations demonstrate that mechanical twist can be used to modify the transmission signatures of interacting photon modes, suggesting a potential route towards microwave devices with mechanically switchable level-attraction-like and level-repulsion-like responses.

\section*{Acknowledgements}

This research was supported by the Australian Research Council Centre of Excellence for Dark Matter Particle Physics under Grant No. CE200100008. Emma Paterson also acknowledges partial financial support from the Defence Science Centre, an initiative of the State Government of Western Australia.

\end{document}